\documentclass[nofootinbib,superscriptaddress,longbibliography,aps,prd,reprint,amsmath,amssymb]{revtex4-2}
\usepackage{bm}
\usepackage{graphicx}
\usepackage{hyperref}
\usepackage{siunitx}

\newcommand*{\rmi}{\mathrm{i}}
\newcommand*{\rmd}{\mathrm{d}}
\newcommand*{\ini}{\mathrm{ini}}

\newcommand*{\wpr}{\omega_\mathrm{P}}

\newcommand*{\sfH}{\mathsf{H}}
\newcommand*{\sfC}{\mathsf{C}}

\newcommand*{\bfB}{\mathbf{B}}
\newcommand*{\bfP}{\mathbf{P}}
\newcommand*{\bbP}{\bar{\mathbf{P}}}
\newcommand*{\bfD}{\mathbf{D}}

\newcommand*{\bP}{\bar{P}}
\newcommand*{\bS}{\bar{S}}
\newcommand*{\bGamma}{\bar{\Gamma}}

\newcommand*{\figscale}{0.58}

\newcommand*{\pfrac}[2]{\left(\frac{#1}{#2}\right)}
\begin{document}

\title{Collision-induced flavor instability in dense neutrino gases with energy-dependent scattering}

\newcommand{\UNM}{Department of Physics and Astronomy, University of New Mexico, Albuquerque, New Mexico 87131, USA}
\newcommand{\ASUastro}{Department of Astronomy/Steward Observatory, University of Arizona, Tucson, Arizona 85721, USA}
\newcommand{\ASUphys}{Department of Physics, University of Arizona, Tucson, Arizona 85721, USA}

% \author[1,2,3]{Yu-Chia Lin}
\author{Yu-Chia Lin}
\email{yuchialin@arizona.edu}
\affiliation{\UNM}
\affiliation{\ASUastro}
\affiliation{\ASUphys}
% \altaffiliation{Current address: Department of Astronomy and Department of Physics, University of Arizona, Tuscon, Arizona, 85721, USA}
% \author[1]{Huaiyu Duan\corref{cor1}}%\,\orcidlink{0000-0001-6708-3048}}
\author{Huaiyu Duan}
\email{duan@unm.edu}
\affiliation{\UNM}
% \affiliation{Department of Physics \& Astronomy, University of New Mexico, Albuquerque, New Mexico 87131, USA}

% \cortext[cor1]{Corresponding author}
% \fntext[fn1]{Current address: Department of Astronomy and Department of Physics, University of Arizona, Tuscon, Arizona, 85721, USA}
% \address{Department of Physics \& Astronomy, University of New
%   Mexico, Albuquerque, New Mexico 87131, USA}
% \affiliation[1]{organization={Department of Physics \& Astronomy, University of New Mexico}, city={Albuquerque}, state={New Mexico}, postcode={87131}, country={USA}}
% \affiliation[2]{organization={Department of Astronomy, University of Arizona}, city={Tucson}, state={Arizona}, postcode={85721}, country={USA}}
% \affiliation[3]{organization={Department of Physics, University of Arizona}, city={Tucson}, state={Arizona}, postcode={85721}, country={USA}}

\date{\today}

\begin{abstract}
    We investigate the collision-induced flavor instability in homogeneous, isotropic, dense neutrino gases in the two-flavor mixing scenario with energy-dependent scattering. We uncover a simple expression of the growth rate of this instability in terms of the flavor-decohering collision rates and the electron lepton number distribution of the neutrino. This growth rate is common to the neutrinos and antineutrinos of different energies, and is independent of the mass-splitting and vacuum mixing angle of the neutrino, the matter density, and the neutrino density, although the initial amplitude of the unstable oscillation mode can be suppressed by a large matter density. Our results suggest that neutrinos are likely to experience collision-induced flavor conversions deep inside a core-collapse supernova even when both the fast and slow collective flavor oscillations are suppressed. 
    %However, this phenomenon may not occur in a neutron star merger because the electron antineutrinos have larger average energy and more abundance than the electron neutrinos in such an environment. 
\end{abstract}

% \begin{keyword}
%     neutrino oscillation \sep dense neutrino medium \sep core-collapse supernova
% \end{keyword}

\maketitle
\section{Introduction}

Neutrinos help shape the physical and chemical evolution of the early universe, core-collapse supernovae, and neutron star mergers where they are copiously produced. Flavor oscillations, which alter the flavor composition of the neutrinos (see, e.g., Ref.~\cite{Workman:2022ynf} for a review), can have a significant impact on the physical conditions in these interesting astrophysical environments. In addition to the well-known vacuum oscillations \cite{Pontecorvo:1967fh, Gribov:1968kq} and the Mikheyev-Smirnov-Wolfenstein (MSW) effect \cite{Wolfenstein:1977ue, Mikheyev:1985zog}, the ambient neutrinos further change the refraction of the neutrinos in these extreme environments \cite{Fuller:1987aa,Notzold:1987ik,Pantaleone:1992xh}. As a result, the dense neutrino gas can experience a collective flavor transformation because of the tight coupling among the neutrinos themselves \cite{Samuel:1993uw,Pastor:2002we,Duan:2006jv}. (See, e.g., Ref.~\cite{Duan:2010bg,Chakraborty:2016yeg} for reviews on this topic and the references therein.) It has been shown that the so-called fast flavor conversions, a special type of collective flavor transformation that arises on the scales of centimeters to meters, can take place near or even below the neutrino-decoupling layer of a neutrino-emission compact object \cite{Sawyer:2015dsa,Chakraborty:2016lct}. (See also Ref.~\cite{Tamborra:2020cul} for a review and the references therein.) An interesting recent development in the research of collective neutrino oscillations is that neutrino collisions, which are usually thought to damp neutrino oscillations, are shown to be able to induce flavor conversions \cite{Capozzi:2018clo,Shalgar:2020wcx,Sasaki:2021zld,Johns:2021qby,Hansen:2022xza,Johns:2022bmu,Kato:2022vsu,Johns:2022yqy,Padilla-Gay:2022wck,Xiong:2022vsy}.

In this work, we focus on the collision-induced flavor conversions in homogeneous and isotropic neutrino gases. We first follow the pioneering work of Johns \cite{Johns:2021qby} and solve the flavor evolution in a mono-energetic neutrino gas. By comparing the numerical and analytical solutions, we demonstrate the dependence of the collision-induced instability on the effective mixing angle and the density of the neutrino (Sec.~\ref{sec:mono}). We then consider the effect of energy-dependent neutrino scattering and derive a simple expression for the exponential growth rate of the collision-induced flavor instability for which we also provide a few illustrative numerical examples (Sec.~\ref{sec:ces}). We conclude by summarizing our results and discussing their implications (Sec.~\ref{sec:conclusions}). 

\section{Mono-energetic neutrino gas}\label{sec:mono}
\subsection{Physics model}
The flavor content of a dense neutrino gas can be represented by the flavor density matrix $\rho$ whose diagonal elements are the neutrino occupancies in different weak-interaction states and the off-diagonal elements are the coherences among those states \cite{Sigl:1993ctk}. For a homogeneous and isotropic environment, the flavor evolution of the neutrino gas can be solved from the following equation:
\begin{align}
    \dot{\rho} = -\rmi[\sfH, \rho] + \sfC,
    \label{eq:eom-rho}
\end{align}
where $\sfH$ and $\sfC$ are the flavor-evolution Hamiltonian and the collision term for the neutrino, respectively.
In the two-flavor mixing scenario, say between the $e$ and $x$ flavors, one can expand $\rho$ in terms of the $2\times2$ identity matrix $\sigma_0$ and the Pauli matrices $\sigma_i$ ($i=1,2,3$) so that
\begin{align}
    \rho \propto \sigma_0 P_0 + \bfP\cdot\boldsymbol{\sigma},
\end{align}
where $\bfP$ is the flavor polarization (Bloch) vector, and $P_0$ is proportional to the total density of the $e$ and $x$ flavor neutrinos. The corresponding quantities $\bar\rho$, $\bP_0$, and $\bbP$ can be defined for the antineutrinos in a similar way.

We first consider a mono-energetic neutrino gas of a single vacuum oscillation frequency $\omega=\Delta m^2/2E$, where $\Delta m^2$ and $E$ are the mass-squared difference and energy of the neutrino, respectively. We assume a minimal collisional model that damps the flavor coherence without changing the particle numbers \cite{Raffelt:1992uj}. In this model, Eq.~\eqref{eq:eom-rho} simplifies as 
\begin{subequations}
    \label{eq:eom}
    \begin{align}
        \dot{\bfP} &= \omega\bfB\times\bfP + \mu\bfD\times\bfP 
        - \Gamma\bfP^\perp,  \\
        \dot{\bbP} &= -\omega\bfB\times\bbP + \mu\bfD\times\bbP 
        - \bGamma\bbP^\perp,  
    \end{align}
\end{subequations} 
where $\bfB=(s_{2\theta}, 0, -c_{2\theta})$ in the flavor basis with $s_{2\theta}=\sin(2\theta)$ and $c_{2\theta}=\cos(2\theta)$, respectively, $\mu=\sqrt2G_\mathrm{F}n_\nu^0$ is the strength of the neutrino self-coupling potential, $\bfD=\bfP-\bbP$, $\bfP^\perp=(P_1, P_2, 0)$ and $\bbP^\perp=(\bP_1, \bP_2, 0)$ represent the flavor coherences of the neutrino and antineutrino, respectively, and $\Gamma$ and $\bGamma$ are the corresponding flavor-decohering neutrino collision rates.%
% \footnote{%
% %The dominant flavor-decohering processes in core-collapse supernovae and binary neutron star mergers are $\nu_e + n \rightleftharpoons p + e^-$ and $\bar\nu_e + p \rightleftharpoons n + e^+$. These charged-current processes, when occurs to a single neutrino, ``measures'' its flavor and thus damps the flavor oscillation. 
% In addition to the flavor-decohering effect, Ref.~\cite{Johns:2021qby} also includes a term that changes the numbers of $\nu_e$ and $\bar\nu_e$ due to the emission and absorption of these neutrinos. This additional term, when included, brings $\nu_e$ and $\bar\nu_e$ back to the equilibrium values after the collision-induced flavor conversion subsides.}
As in Ref.~\cite{Johns:2021qby}, we approximate the matter suppression on collective neutrino oscillations by a small effective mixing angle $\theta$ \cite{Hannestad:2006nj}. We also choose a nominal neutrino density $n_\nu^0$ to make $\bfP$ and $\bbP$ dimensionless.  We focus on the physical environments where both the matter density and the neutrino density are so large that $\theta$, $\omega/\mu$, $\Gamma/\mu$, and $\bGamma/\mu$ are all much less than 1.

\begin{figure}[htb]
    \includegraphics[trim=1 2 1 1, clip, scale=\figscale]{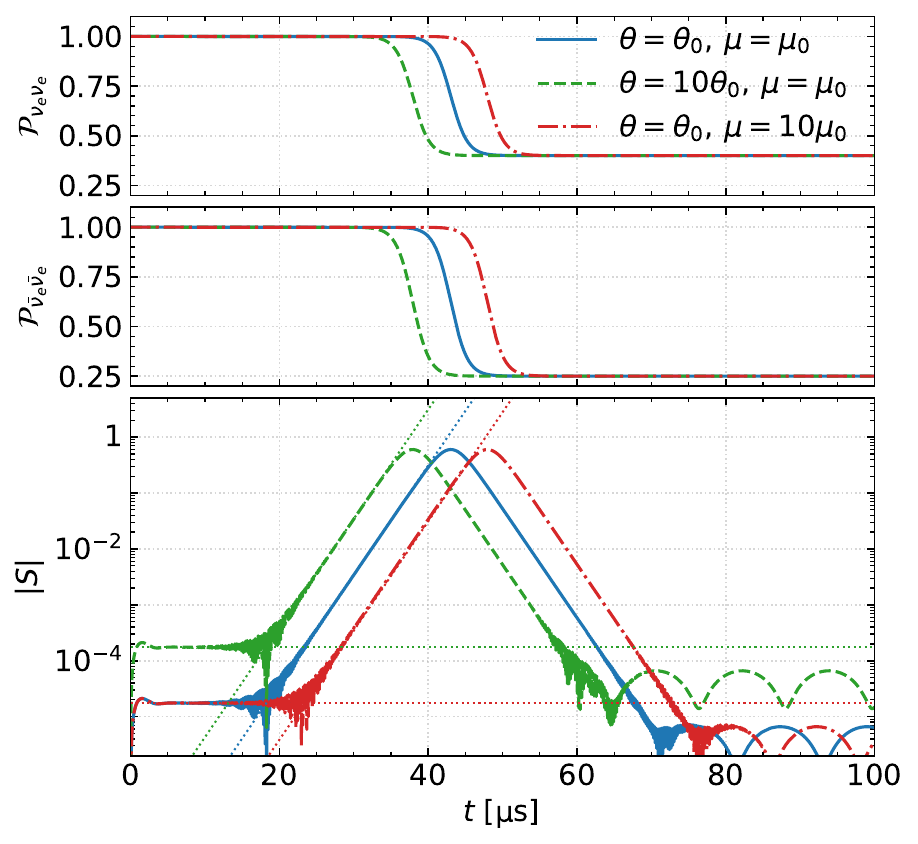}
    \caption{The survival probabilities of the electron flavor neutrino (upper panel) and antineutrino (middle panel) and the flavor coherence of the neutrino (bottom panel) as functions of time in the mono-energetic neutrino gases with three combinations of the effective mixing angle $\theta$ and the neutrino self-coupling strength $\mu$ (as labeled), where $\theta_0=10^{-5}$ and $\mu_0=\SI{e5}{km^{-1}}$. The horizontal dotted lines in the bottom panel represent $|Q_{0,1}|$ in these scenarios (with two of them overlapping with each other), and the slanted dotted lines are $|Q_{+,1} e^{-\rmi\Omega_+ t}|$ [see Eq.~\eqref{eq:1E-sol}].}
    \label{fig:PS-t}
\end{figure}

As concrete examples, we solve Eq.~\eqref{eq:eom} numerically with the initial conditions $\bfP(t=0)=\bfP^\ini=(0, 0, 1)$ and $\bbP^\ini=(0,0,0.8)$ and with $\omega=\SI{0.6}{km^{-1}}$, $\Gamma=\SI{1}{km^{-1}}$, $\bGamma=\SI{0.5}{km^{-1}}$, and three combinations of $\theta$ and $\mu$: ($\theta_0$, $\mu_0$), ($10\theta_0$, $\mu_0$) and ($\theta_0$, 10$\mu_0$), where $\theta_0=10^{-5}$ and $\mu_0=\SI{e5}{km^{-1}}$, respectively. We plot in Fig.~\ref{fig:PS-t} the survival probabilities of $\nu_e$ and $\bar\nu_e$, which are computed as
\begin{align}
    \mathcal{P}_{\nu_e\nu_e} = \frac{1+P_3/P^\ini_3}{2}
    \quad\text{and}\quad
    \mathcal{P}_{\bar\nu_e\bar\nu_e} = \frac{1+\bP_3/\bP^\ini_3}{2},
    \label{eq:Psur}
\end{align}
respectively, as well as the magnitude of the flavor coherence
\begin{align}
    S = P_1 -\rmi P_2
    \label{eq:S}
\end{align}
of the neutrino.

In all three cases, the neutrino gases experience flavor conversions $\nu_e\rightarrow\nu_x$ and $\bar\nu_e\rightarrow\bar\nu_x$ between \SI{35}{\us} and \SI{50}{\us}. While the gases start with an excess of $\nu_e$ and $\bar\nu_e$, they end up with more $\nu_x$ and $\bar\nu_x$.
The exponential growth of $|S|$ between \SI{20}{\us} and \SI{40}{\us} confirms that the flavor conversions are indeed due to some flavor instability \cite{Johns:2021qby}. The growth rate of $|S|$ is largely independent of $\theta$ or $\mu$ when $S$ is small. However, the flavor conversions are delayed by a smaller $\theta$ and/or a larger value of $\mu$, which represent a larger matter density and a larger neutrino density, respectively. 

Although we plot the evolution of the neutrino gases up to \SI{100}{\us} in Fig.~\ref{fig:PS-t} for the completeness of the solution, one should note Eq.~\eqref{eq:eom} is valid only in the linear regime. Additional terms that change the populations of different neutrino species must be included for the complete treatment in the nonlinear regime. When included, these terms bring $\nu_e$ and $\bar\nu_e$ back to the equilibrium values after the collision-induced flavor conversion subsides \cite{Johns:2021qby}.

\subsection{Collision-induced flavor instability}

To understand the dependence of the collision-induced flavor instability on $\theta$ and $\mu$, we linearize Eq.~\eqref{eq:eom} when $S$ and $\bS$ are small \cite{Johns:2021qby,Padilla-Gay:2022wck}:
\begin{align}
    \rmi\frac{\rmd}{\rmd t}
    \begin{bmatrix}
        S \\ \bS
    \end{bmatrix}
    \approx \omega s_{2\theta} \begin{bmatrix}
        -P^\ini_3 \\ \bP^\ini_3
    \end{bmatrix}
    + \Lambda
    \begin{bmatrix}
        S \\ \bS
    \end{bmatrix},
    \label{eq:lin-eom}
\end{align}
where 
\begin{align}
    \Lambda = \begin{bmatrix}
        -\omega c_{2\theta} - \mu \bP^\ini_3 -\rmi\Gamma & \mu P^\ini_3 \\
        -\mu \bP^\ini_3 & \omega c_{2\theta} +\mu P^\ini_3 -\rmi\bGamma
    \end{bmatrix}.
\end{align}
Equation~\eqref{eq:lin-eom} has the solution
\begin{align}
    \begin{bmatrix}
        S(t) \\ \bS(t)
    \end{bmatrix}
    = Q_0
    + Q_+ e^{-\rmi\Omega_+ t}
    + Q_- e^{-\rmi\Omega_- t},
\end{align}
where
\begin{align}
    Q_0 = -\omega s_{2\theta} \Lambda^{-1} 
    \begin{bmatrix}
        -P^\ini_3 \\ \bP^\ini_3
    \end{bmatrix},
\end{align}
while $\Omega_\pm$ and $Q_\pm$ are the eigenvalues and eigenvectors of $\Lambda$, respectively. The amplitudes of $Q_\pm$ are constrained by the initial condition
\begin{align}
    0 = \begin{bmatrix}
        S(0) \\ \bS(0)
    \end{bmatrix}
    = Q_0 + Q_+ + Q_- 
\end{align}
or, equivalently,
\begin{align}
    \omega s_{2\theta} \begin{bmatrix}
        -P^\ini_3 \\ \bP^\ini_3
    \end{bmatrix}
    = \Omega_+ Q_+ + \Omega_- Q_-.
\end{align}
In the lowest non-vanishing orders of $\theta$, $\omega/\mu$, $\Gamma/\mu$, and $\bGamma/\mu$, we find
\begin{subequations}
    \label{eq:1E-sol}
    \begin{align}
        \Omega_+ &\approx \mu D_3 - \rmi\pfrac{P^\ini_3\bGamma - \bP^\ini_3\Gamma}{D_3}, 
        \label{eq:Omega-plus}\\
        \Omega_- &\approx -\omega\pfrac{P^\ini_3 + \bP^\ini_3}{D_3} - \rmi\pfrac{P^\ini_3\Gamma - \bP^\ini_3\bGamma}{D_3}, 
        \label{eq:Omega-minus}\\
        Q_0 &\approx \frac{-2\omega\theta(P^\ini_3+\bP^\ini_3)}{\omega(P^\ini_3+\bP^\ini_3)+\rmi(\Gamma P^\ini_3 - \bGamma\bP^\ini_3)}\begin{bmatrix}
            P^\ini_3 \\ \bP^\ini_3
        \end{bmatrix} ,\\
        Q_+ &\approx \frac{4\omega\theta P^\ini_3  \bP^\ini_3}{\Omega_+ D_3}
        \begin{bmatrix}
            1 \\ 1
        \end{bmatrix}, \\
        Q_- &\approx -\frac{2\omega\theta(P^\ini_3 + \bP^\ini_3)}{\Omega_- D_3}
        \begin{bmatrix}
            P^\ini_3 \\ \bP^\ini_3
        \end{bmatrix},
    \end{align}    
\end{subequations}
where $D_3=P_3-\bP_3$ is approximately a constant of motion for Eq.~\eqref{eq:eom} when $\theta\ll1$. The expressions of $\Omega_\pm$ in the above equation agree with those in Ref.~\cite{Padilla-Gay:2022wck}, and they correspond to the possible two types of flavor instability pointed out in Ref.~\cite{Johns:2021qby}.

We note that $Q_+\propto \theta\omega/\mu$ and $Q_-$ is independent of $\mu$. In the above numerical examples, the plus mode is unstable with the exponential growth rate
\begin{align}
    \gamma_+ \approx \frac{\Gamma\bP^\ini_3 - \bGamma P^\ini_3}{D_3},
\end{align}
and the minus mode has a exponential damping rate
\begin{align}
    \gamma_- \approx -\frac{\Gamma P^\ini_3 - \bGamma \bP^\ini_3}{D_3}.
    \label{eq:gamma-}
\end{align}
The dependence of $Q_+$ on $\theta$ and $\mu$ explains the apparent delay of the flavor conversions for a small effective mixing angle and/or a large neutrino density. This dependence also implies that a very tight error tolerance is required to solve Eq.~\eqref{eq:eom} numerically when both $\theta$ and $\omega/\mu$ are small.

We have chosen $\bGamma<\Gamma$ in the above examples. However, the electron antineutrinos tend to have larger average energy than the neutrinos and, therefore, can have a larger effective collision rate $\bGamma$. As we will see in the next section,  the minus mode can become unstable when the energy dependence of the collision rates is taken into account.

\section{Neutrino gas with a continuous energy spectrum}\label{sec:ces}
\subsection{Linear stability analysis}
For a homogeneous, isotropic neutrino gas with a continuous energy spectrum, Eq.~\eqref{eq:eom} is generalized to:
\begin{align}
    \dot{\bfP}_\omega &= (\omega\bfB + \mathbf{L} + \mu\bfD)\times\bfP_\omega 
        - \Gamma_\omega\bfP^\perp_\omega.
    \label{eq:eom-mE}
\end{align}
Here we have restored the matter potential $\mathbf{L}=(0,0,\lambda) = (0,0,\sqrt{2}G_\mathrm{F}n_e)$, where $n_e$ is the net electron density. We adopt the flavor isospin notation \cite{Duan:2005cp} with
\begin{align}
    \bbP_\omega = -\bfP_{-\omega}
\end{align}
so that
\begin{align}
    \bfD = \int_{-\infty}^\infty \bfP_\omega\,\rmd\omega.
\end{align}
The flavor instability of the neutrino gas can be obtained by linearizing Eq.~\eqref{eq:eom-mE} when $|S_\omega|\ll1$ \cite{Banerjee:2011fj}:
\begin{align}
    \rmi\dot{S}_\omega = -(\omega c_{2\theta} + \rmi\Gamma_\omega) S_\omega - \mu g(\omega) \int_{-\infty}^\infty S_{\omega'}\,\rmd\omega',
    \label{eq:eom-lin-mE}
\end{align}
where 
\begin{align}
    g(\omega)=P^\ini_{\omega,3}
\end{align}
is the initial neutrino electron lepton number ($\nu$ELN) distribution or the flavor(-difference) distribution of the neutrino. The total $\nu$ELN 
\begin{align}
    D_3 \approx \int_{-\infty}^\infty g(\omega)\,\rmd\omega
\end{align}
is approximately a constant of motion for Eq.~\eqref{eq:eom-mE} when both $\lambda$ and $\mu$ are much larger than $\omega$.
As usual, we have dropped a term in Eq.~\eqref{eq:eom-lin-mE} which is proportional to $\lambda + \mu D_3$ by choosing an appropriate co-rotating frame \cite{Duan:2005cp}. Applying the collective oscillation ansatz $S_\omega(t) = Q_\omega e^{-\rmi\Omega t}$ to Eq.~\eqref{eq:eom-lin-mE}, we obtain
\begin{align}
    \Omega Q_\omega = -(\omega c_{2\theta} + \rmi\Gamma_\omega) Q_\omega - \mu C g(\omega)
\end{align} 
or
\begin{align}
    Q_\omega = - \frac{\mu C g(\omega) }{\Omega + \omega c_{2\theta} +\rmi\Gamma_\omega},
    \label{eq:Q}
\end{align}
where
\begin{align}
    C = \int_{-\infty}^\infty Q_\omega\,\rmd\omega.
    \label{eq:Q-int}
\end{align}
Substituting Eq.~\eqref{eq:Q} into Eq.~\eqref{eq:Q-int} we obtain
\begin{align}
    \int_{-\infty}^\infty\frac{g(\omega)\,\rmd\omega}{\Omega+\omega c_{2\theta}+ \rmi\Gamma_\omega}=-\frac{1}{\mu}
    \label{eq:self-cons-1}
\end{align}
or
\begin{subequations}
    \label{eq:self-cons}
    \begin{align}
        \int_{-\infty}^\infty\frac{g(\omega)(\wpr+\omega c_{2\theta})\,\rmd\omega}{(\wpr+\omega c_{2\theta})^2 + (\gamma+\Gamma_\omega)^2}&=-\frac{1}{\mu}, 
        \intertext{and}
        \int_{-\infty}^\infty\frac{g(\omega)(\gamma+\Gamma_\omega)\,\rmd\omega}{(\wpr+\omega c_{2\theta})^2 + (\gamma+\Gamma_\omega)^2}&=0, 
    \end{align}        
\end{subequations}
where $\Omega = \wpr + \rmi\gamma$. Because $\Omega$ is the only quantity in the left-hand side of Eq.~\eqref{eq:self-cons-1} that depends on $\mu$, and because $\gamma$ is expected to be of the same order as typical $\Gamma_\omega$, we assume $\wpr\propto\mu$ in the large $\mu$ limit and obtain the following solution for Eq.~\eqref{eq:self-cons}:%
% \footnote{This analysis is not valid when $D_3\approx0$, a scenario will be addressed in a follow-up work.}
\begin{align}
    \wpr \approx -\mu D_3
\end{align}
and
% \begin{align}
%     \int_{-\infty}^\infty g(\omega)(\gamma + \Gamma_\omega)\,\rmd\omega \approx 0
% \end{align}
% from which we obtain
\begin{align}
    \gamma \approx -\frac{1}{D_3}\int_{-\infty}^\infty g(\omega)\Gamma_\omega\,\rmd\omega.
    \label{eq:gamma}
\end{align}
In other words, the exponential growth rate of the collision-induced flavor instability, when it exists, is the negative average of the flavor-decohering collision rates of the neutrinos weighted by the $\nu$ELN distribution. 

To make connection with the mono-energetic case, we define 
\begin{align}
    \Gamma = \frac{\int_0^\infty g(\omega)\Gamma_\omega\,\rmd\omega}{\int_0^\infty g(\omega)\,\rmd\omega} 
    \quad\text{and}\quad
    \bGamma = \frac{\int_{-\infty}^0 g(\omega)\Gamma_\omega\,\rmd\omega}{\int^0_{-\infty} g(\omega)\,\rmd\omega} 
\end{align}
as the effective collision rates for the neutrinos and the antineutrinos, respectively, which are represented by the total polarization vectors
\begin{align}
    \bfP = \int_0^\infty \bfP_\omega \,\rmd\omega
    \quad\text{and}\quad
    \bbP = -\int^0_{-\infty} \bfP_\omega\,\rmd\omega.
\end{align}
With these definitions, Eq.~\eqref{eq:gamma} bears the same form as Eq.~\eqref{eq:gamma-} and corresponds to the minus mode in the mono-energetic limit.

The mono-energetic limit provides an interesting explanation for the seemingly counter-intuitive phenomenon that neutrino collisions can lead to flavor instability in a dense neutrino gas. It is true that neutrino collisions tend to reduce the flavor coherences of both the neutrinos and antineutrinos which are represented by the magnitudes of $\bfP^\perp$ and $\bbP^\perp$, respectively. When the effective collision rate of $\bar\nu_e$ is sufficiently larger than that of $\nu_e$, $\bbP^\perp$ shrinks faster than $\bfP^\perp$ which results in a net increase of $\bfD^\perp=\bfP^\perp-\bbP^\perp$ when there are more $\nu_e$ than $\bar\nu_e$ initially. In a very dense neutrino gas, both $\bfP$ and $\bbP$ are locked in the anti-aligned orientations and their evolution is mainly driven by $\bfD=\bfP-\bbP$ [see Eq.~\eqref{eq:eom}]. An increase of $\bfD^\perp$ implies a further tilt away of $\bfD$ and, therefore, both $\bfP$ and $\bbP$ from their original directions, which is the cause of the flavor instability.

Equation~\eqref{eq:gamma} shows that the existence of the collision-induced flavor instability requires a crossing in the $\nu$ELN distribution $g(\omega)$ because the neutrino collision rates $\Gamma_\omega$ are always positive. This is reminiscent of a similar condition in the case of a collisionless neutrino gas \cite{Dasgupta:2009mg} and is a special case of the general requirement for the flavor instability \cite{Dasgupta:2021gfs}. In addition, the growth rate of this instability depends only on the shape of the $\nu$ELN distribution $g(\omega)$ and the decohering neutrino collision rates $\Gamma_\omega$. It is independent of the matter density, the neutrino density, and the neutrino mass-splitting when both $\mu$ and $\lambda$ are large.

\subsection{Numerical examples}

\begin{figure}[htb]
    \includegraphics[trim=1 2 1 1, clip, scale=\figscale]{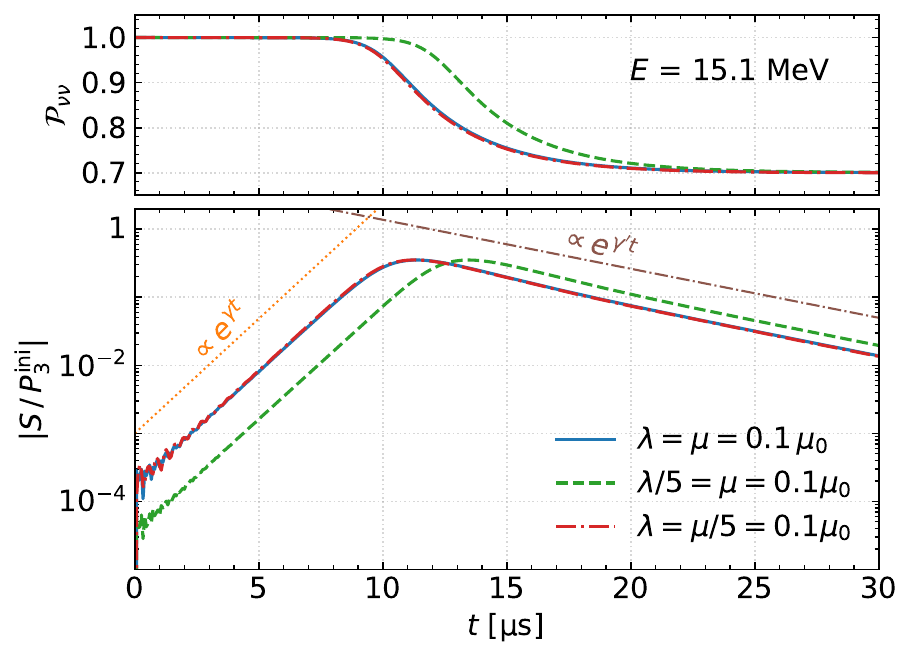}
    \caption{Similar to Fig.~\ref{fig:PS-t} but for the \SI{15.1}{MeV} neutrino in the neutrino gas with a continuous energy spectrum. The results for three combinations of the strength of the matter potential $\lambda$ and that of the neutrino potential $\mu$ (as labeled) are shown as the solid, dashed, and dot-dashed curves, respectively, two of which largely overlap with each other. The survival probabilities of the neutrino and the antineutrino are virtually the same and are not plotted separately. The slanted dotted line in the bottom panel represents a pure exponential growing function $\sim e^{\gamma t}$ with $\gamma\approx \SI{2.59}{km^{-1}}$ predicted by Eq.~\eqref{eq:gamma}. The slanted dot-dashed line represents an exponential decay function $\sim e^{\gamma' t}$ with $\gamma'\approx\SI{-0.553}{km^{-1}}$ which is determined by Eq.~\eqref{eq:gamma} with the final $\nu$ELN distribution.}
    \label{fig:PS-t-mE}
\end{figure}

As illustrative examples, we consider the Fermi-Dirac distribution for the initial neutrino energy spectrum with
\begin{align}
    g(\omega) \propto \begin{cases}
        E^4/[\exp(E/T_{\nu_e}) + 1] & \text{ if } \omega > 0, \\
        -E^4/[\exp(E/T_{\bar\nu_e}) + 1] & \text{ if } \omega < 0,
    \end{cases}
\end{align}
where $E(\omega)=|\Delta m^2/2\omega|$, and the extra factor of $E^2$ stems from $\rmd E/\rmd\omega$. We normalize the initial $\nu$ELN distribution by the following conditions:
\begin{align}
    \int_0^\infty g(\omega)\,\rmd\omega = 1
    \quad\text{and}\quad
    \int_{-\infty}^0 g(\omega)\,\rmd\omega = -0.8.
\end{align}
We choose $T_{\nu_e}=\SI{4}{MeV}$ and $T_{\bar\nu_e}=\SI{5}{MeV}$ which give the initial average energies $\langle E_{\nu_e}\rangle \approx \SI{12.6}{MeV}$ and $\langle E_{\bar\nu_e}\rangle \approx \SI{15.8}{MeV}$, respectively. We discretize the energy spectrum of the neutrino (as well as that of the antineutrino) into 400 equal-sized bins between 0 and 80 MeV and solve Eq.~\eqref{eq:eom-mE} numerically for three combinations of $\lambda$ and $\mu$: $(\mu_0/10, \mu_0/10)$, $(\mu_0/2, \mu_0/10)$, and $(\mu_0/10, \mu_0/2)$ with $\mu_0=\SI{e5}{km^{-1}}$. We also choose $\Delta m^2=\Delta m_\mathrm{atm}^2\approx \SI{2.5e-3}{eV^{-2}}$, $\theta=\theta_{13}\approx0.15$, and%
\footnote{For the neutrinos with energies much less than \SI{1}{GeV}, their rates of collision with nucleons are proportional to $E^2$. Interested readers can refer to, e.g., Ref.~\cite{Johns:2021qby} for an estimate of these rates.}
\begin{align}
    \Gamma_\omega = \SI{1}{km^{-1}}\pfrac{E}{\SI{10}{MeV}}^2.
\end{align}

\begin{figure}[htb]
    \includegraphics[trim=1 2 1 1, clip, scale=\figscale]{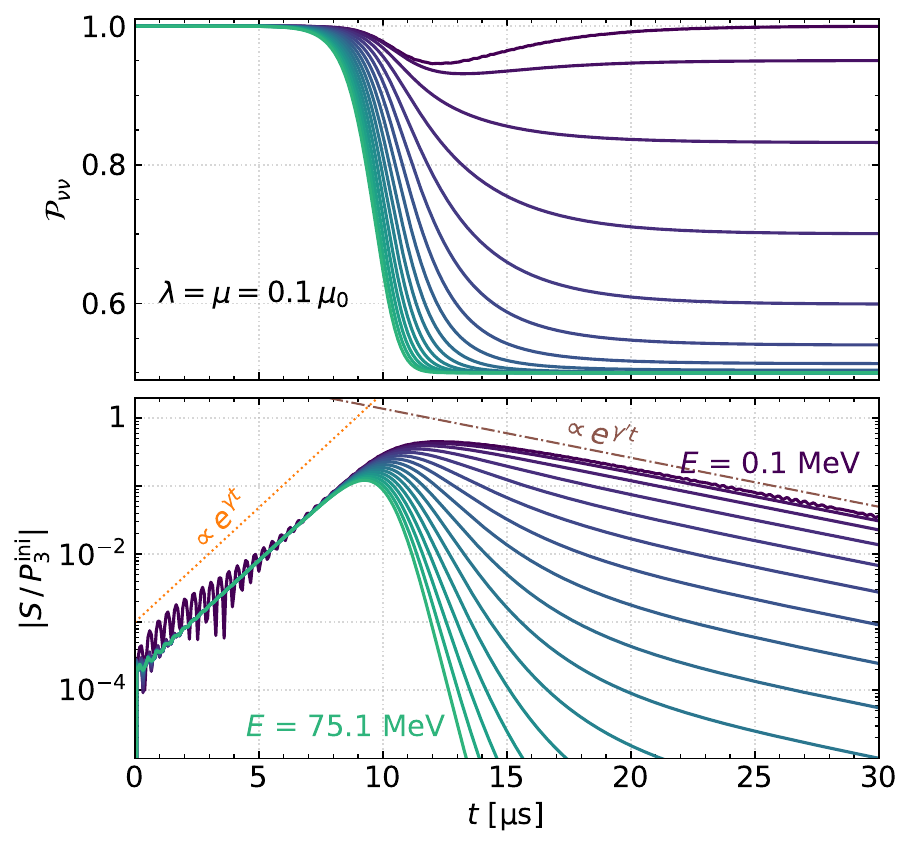}
    \caption{Similar to Fig.~\ref{fig:PS-t-mE} but for the case $\mu=\lambda=0.1\mu_0$ only and for the neutrinos with energies from \SI{0.1}{MeV} (darkest solid curves) to \SI{75.1}{MeV} (lightest solid curves) with an increment of 5 MeV (as labeled).} 
    \label{fig:PS-t-mE-fid}
\end{figure}

In Fig~\ref{fig:PS-t-mE} we plot the neutrino survival probability $\mathcal{P}_{\nu\nu}$ and the flavor coherence $S$ of the \SI{15.1}{MeV} neutrino in all three cases. We also plot these quantities for various neutrino energies in the case with $\lambda=\mu=\mu_0/10$ in Fig.~\ref{fig:PS-t-mE-fid}. We note that, similar to fast flavor conversions \cite{Sawyer:2015dsa,Chakraborty:2016lct}, the flavor conversions in these neutrino gases are independent of $\Delta m^2$ when $|\omega|\ll\mu$. Unlike typical fast flavor conversions, however, collision-induced flavor conversions can depend on the neutrino energy through the energy-dependent collision rates $\Gamma_\omega$. Because we assume the same collision rate for $\nu_e$ and $\bar\nu_e$ of the same energy, both $\mathcal{P}_{\nu\nu}$ and $S$ are virtually the same for both the neutrino and the antineutrino.

As in the mono-energetic gases, the flavor coherences of the neutrinos in a dense gas with a continuous energy spectrum experience exponential growth induced by the collision. As a comparison, we plot in the lower panels of Figs.~\ref{fig:PS-t-mE} and \ref{fig:PS-t-mE-fid} the exponential function $e^{\gamma t}/1000$ as the dotted line, where $\gamma\approx \SI{2.59}{km^{-1}}$ is solved from Eq.~\eqref{eq:gamma}. This exponential growth rate is independent of the matter density or the neutrino density (Fig.~\ref{fig:PS-t-mE}). Increasing the matter density has a similar effect as decreasing the effective mixing angle in the mono-energetic case both of which decrease the initial flavor coherences. Unlike the mono-energetic case, however, increasing the neutrino density has little impact on the initial flavor coherences. This shows that the collective unstable mode here is like the minus mode in the mono-energetic case where the initial amplitude is also independent of $\mu$ [see Eq.~\eqref{eq:1E-sol}].

Fig.~\ref{fig:PS-t-mE-fid} shows that, although all the neutrinos of different energies behave similarly when the flavor coherences are small, this unison seems to fall apart in the nonlinear regime. The neutrinos of the lowest energies experience only a slight flavor conversion between \SI{10}{\us} and $\SI{20}{\us}$ and are restored to their original flavors afterwards. The neutrinos of the highest energies, however, approach a complete flavor depolarization which is represented by $\mathcal{P}_{\nu\nu}=1/2$. The flavor coherences of most of the neutrinos, however, do fall off at the same rate $\gamma'$ which is also determined by Eq.~\eqref{eq:gamma} except with $g(\omega)$ replaced by the final $\nu$ELN distribution 
\begin{align}
    g^\mathrm{fin}(\omega) = \lim_{t\rightarrow\infty} P_{\omega,3}(t).
\end{align}

\begin{figure}[htb]
    \includegraphics[trim=1 2 1 1, clip, scale=\figscale]{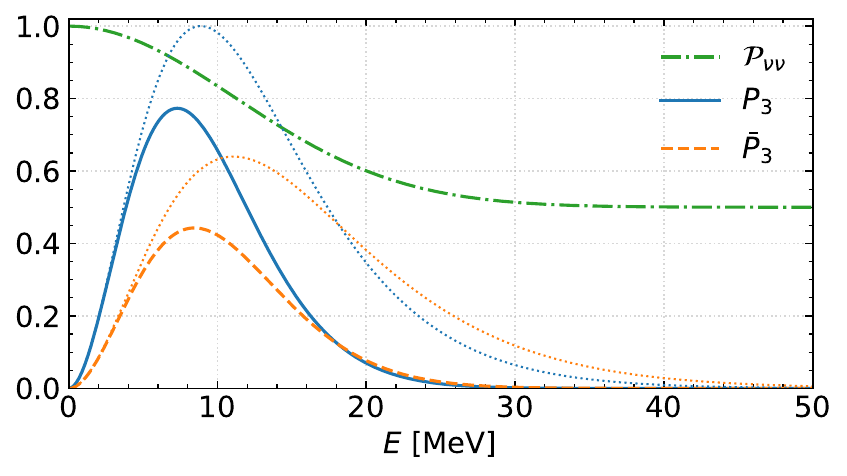}
    \caption{The final neutrino survival probability $\mathcal{P}_{\nu\nu}$ and the polarization vector components $P_3$ and $\bP_3$ as functions of the neutrino energy. The thin dotted curves represent the initial values of $P_3$ and $\bP_3$. The polarization vectors have been rescaled by the maximum value of $P_3$ at $t=0$.}
    \label{fig:P-E-fid}
\end{figure}

In Fig.~\ref{fig:P-E-fid} we plot the final neutrino survival probability $\mathcal{P}_{\nu\nu}(E)$ (at \SI{30}{\us}) as well as the third components of the polarization vectors of both neutrinos and antineutrinos (rescaled by the largest value of $P^\ini_3$). All three test cases with different matter and neutrino densities achieve the same $\mathcal{P}_{\nu\nu}(E)$, $P_3(E)$, and $\bP_3(E)$ in the end.

As in the case of the mono-energetic neutrino gases, we plot the evolution of the neutrino gas beyond the linear regime only for the completeness of the solution. In a real astrophysical environment, however, neutrino emissions and absorptions that change the populations of the neutrino species must be included which are likely to bring the $\nu_e$ and $\bar\nu_e$ densities back to their equilibrium values \cite{Johns:2021qby}. 

\section{Discussion and conclusions} \label{sec:conclusions}

We have analyzed the collision-induced flavor instabilities in the dense, homogeneous and isotropic neutrino gas with a single energy as well as that with a continuous energy spectrum. While there are two collective modes ($\pm$) in the mono-energetic case, it seems that only the minus mode survives in the neutrino gas with a continuous energy spectrum. We showed that the exponential growth rate of the collision-induced collective mode is determined by an average of the flavor-decohering collision rates. [See Eq.~\eqref{eq:gamma}.] This implies that a crossed $\nu$ELN energy distribution is required for the collision-induced collective mode to be unstable. This condition is a special case of the general requirement for the existence of the flavor instability \cite{Dasgupta:2021gfs}. In a very dense neutrino gas, the collision-induced flavor instability is independent of the mass-splitting, the vacuum mixing angle, and the density of the neutrino. A large matter density has no impact on the exponential growth rate of the collective mode either, although it suppresses its initial amplitude as in the case of collisionless flavor instabilities \cite{Hannestad:2006nj}.

Although we focused on the homogeneous and isotropic neutrino gases in this work, collision-induced instabilities can well exist in inhomogeneous and anisotropic environments \cite{Xiong:2022vsy} and even interact with fast flavor conversions under appropriate conditions \cite{Johns:2021qby,Sigl:2021tmj,Shalgar:2022lvv, Kato:2022vsu,Johns:2022yqy,Padilla-Gay:2022wck}. However, it is important to note that collision-induced flavor conversions can occur without the crossing of the angular distribution of the $\nu$ELN which is required for fast flavor instabilities \cite{Izaguirre:2016gsx, Morinaga:2021vmc}. In addition, collision-induced flavor conversions can take place near the surface of or even inside the proto-neutron star where the normal/slow collective flavor oscillations are suppressed by the large matter density and the neutrino density \cite{Hannestad:2006nj,Esteban-Pretel:2008ovd,Duan:2010bf}. We would like to point out that the expression of its exponential growth rate suggests that the collision-induced flavor instability may not exist in the neutron star merger environment. This is because $\bar\nu_e$ has both a larger abundance and a harder spectrum than $\nu_e$ in such an environment (see, e.g., Ref.~\cite{Frensel:2016fge}) which results in an exponential damping of the collective mode. Of course, more detailed research is needed to ascertain the actual impact of collision-induced flavor conversions in real astrophysical environments. For example, a high neutron-to-proton ratio can bump up the effective $\nu_e$ collision rate and changes the picture completely.

\section*{Acknowledgments}
We thank L.~Johns for the useful discussion.
The numerical calculations in this work were performed using \textsc{Julia} \cite{bezanson2017julia,julia} and \textsc{DifferentialEquations.jl} \cite{rackauckas2017differentialequations,differentialequations}.
This work is supported by the US DOE
NP grant No.\ DE-SC0017803 at UNM.

\bibliography{collins}

\end{document}